# Similarity Analysis of Discrete Fracture Networks

Younes Fadakar Alghalandis [a,b,c], Davide Elmo [a], Erik Eberhardt [b]

[a] NBK Institute of Mining Engineering, The University of British Columbia, Vancouver, BC, Canada.
[b] Earth, Ocean & Atmospheric Sciences, The University of British Columbia, Vancouver, BC, Canada.
[c] Alghalandis Computing, New Westminster, BC, Canada.

## Abstract

Applications of Discrete Fracture Network (DFN) modeling are becoming increasingly prevalent in engineering analyses involving fractured rock masses. For example, kinematic evaluations of slope or underground excavation stability and the modeling of fluid flow in fractured rock have been shown to benefit significantly from the explicit representation of DFN realizations in the simulations. In practice, due to high computing costs, namely time, a balance must be struck that limits analyses to the consideration of only a few realizations as input. As a stochastic representation, a single realization is only one possibility. It is therefore critical that the selected realizations (possibilities) are able to summarize the range of variations present in the input parameters adequately for the purpose of study or practice. That is, the significance of diversity (dissimilarity) in the generated fracture networks is of great importance and should be assessed prior to further often time-consuming processing stages. We demonstrate here a novel development in the analysis of the similarity between three-dimensional fracture networks, which provides an accurate, efficient and practical solution with comprehensive coverage of model variations. Several examples are presented together with a comparison between the proposed three-dimensional method and existing methods limited to two-dimensional assumptions. It is shown that the two-dimensional similarity methods despite their popularity are heavily biased and poorly represent the reality.

## 1. Introduction

Discrete Fracture Network (DFN) modeling is increasingly being used in advanced engineering analyses to explicitly account for the presence of discontinuities in fractured rock masses (e.g., Jing 2003, Brady & Brown 2005, Jing & Stephansson 2007, Fadakar-A 2017). This is especially valuable where fractures are expected to govern the behavior of the system, for example where there is a linkage between fracture connectivity and rock slope/underground excavation stability, or hydrocarbon/groundwater flow. In fact, due to recent advancements in computing systems (hardware and software), DFN modeling has established itself as one of the most versatile, powerful and flexible tools for studying complicated rock mass states, responses and interactions related to the extent and complexity of fracture networks (Jing & Stephansson 2007, Fadakar-A 2017). Of value are both the means to analyze the characteristics of the discontinuity network and the evaluation of its response to different processes (Fox et al. 2012). As three-dimensional DFN models are appearing more frequently in the literature (e.g., Erhel et al. 2009, Wettstein et al. 2011, de-Drezy et al. 2013, Kennard et al. 2014, Fadakar-A 2017), the investigations, discussions and outcomes from these studies are helping to promote the development, reliability and value of DFNs to closely represent the true characteristics of the fracture networks to be encountered on projects.

As a stochastic technique, a DFN realization represents only one possibility derived from statistical distributions of dip directions, dip angles, intensities and persistence for each fracture set. It is therefore critical that more than one realization be considered for a detailed engineering analysis and that those implemented are able to capture the range of variations induced by the inherent uncertainties of input parameters. This paper reports the findings of a study discussing the development, implementation and results comparing the similarity between three-dimensional fracture networks, which provides an accurate, efficient and practical solution to assess and achieve comprehensive coverage of model variations. The quantification of the similarity is reported as a single value metric. This is carried out using ADFNE (Fadakar-A 2017), an open source code Matlab package.

*1.1. DFN Similarity and Implications for Rock Engineering Design*

Because of their underlying stochastic nature, there are an infinite number of possible DFN realizations of a three-dimensional fracture system based on the variations present in mapped data. The authors believe that the similarity approach proposed in this paper addresses an important issue that has significant implications for rock engineering design, and the objective of this paper is to introduce an objective similarity metric that could help engineers identify which simulations to run. The Synthetic Rock Mass (SRM) approach is a typical example of integrated DFN-Geomechanical analysis that could greatly benefit from the introduction of a DFN similarity metric. Three main components converge into the SRM approach (Elmo et al. 2016): i) data collection and characterization, ii) DFN modelling, and iii) the geomechanical model used for simulating rock mass behavior. In the literature (e.g. Pierce et al. 2007, Mas-Ivars 2008, Elmo et al. 2012, 2016) authors have typically used a very limited number of DFN realizations to determine a representative range of SRM based rock mass properties for the area/volume under consideration (Fig.1). This limitation arises due to the excessive computing times required to run a single large-scale SRM model. Yet, the determination of the scale at which the SRM properties become representative of the rock mass may require running multiple DFN realizations for SRM models at different scales. For real-life engineering projects the number of DFN realizations that is possible to consider in the analysis may be further constrained by budget considerations. Thus, practitioners are required to limit their analysis to a restricted number of realizations; a specific case is presented in Elmo et al. (2011), in which the authors used three-dimensional DFN models as the source to derive two-dimensional sampling planes to determine the factor of safety of a large rock slope problem (Fig.2). The factor of safety against failure for the slope problem in Fig.2 could be highly sensitive to the location of only a few pre-defined fracture traces. Given the assumption of a biplanar slope failure mode, it is evident that the two-dimensional models are characterized by a relatively wide range of rock bridge lengths (representing increased rock mass cohesive strengths). These would in turn provide considerable additional strength and accordingly increase the modelled factor of safety.

It is argued that similar models would have a similar mechanical behavior (hence a similar factor of safety), so that the availability of an objective similarity metric would provide a valuable means to improve the selection of which two-dimensional models to consider and reduce uncertainty associated with a subjective sampling approach. As a solution and to avoid erroneous and or biased interpretations, the similarity associated decisions should be made based on the three-dimensional similarity metrics such as the one proposed in this paper.

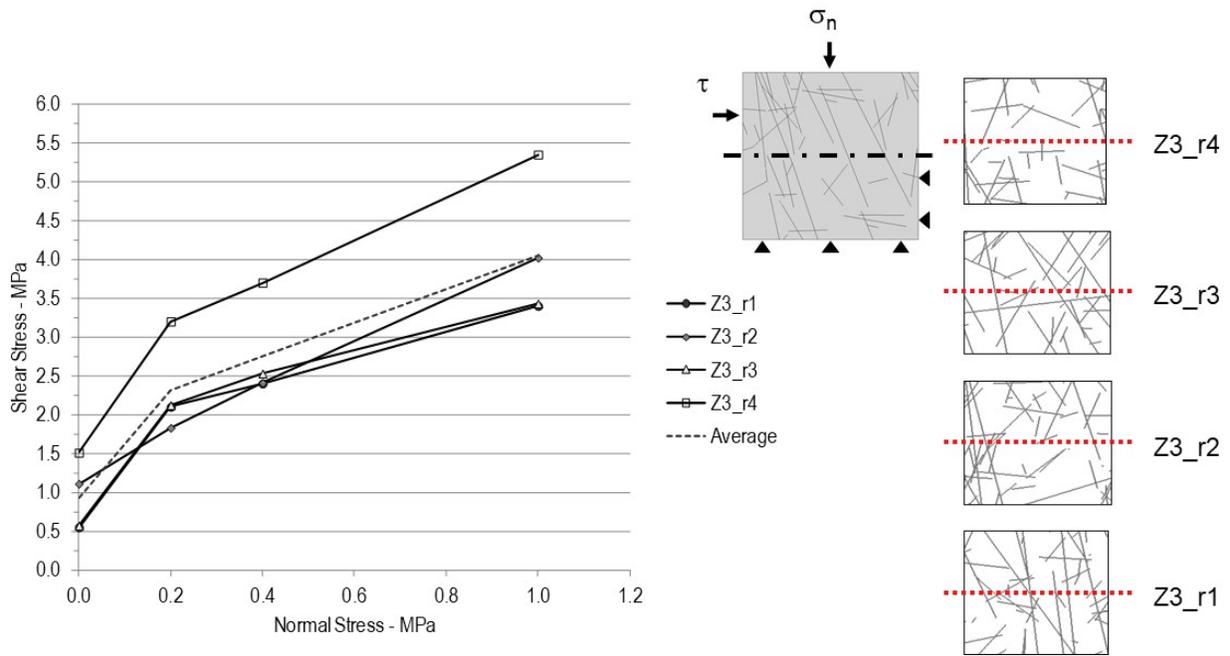

*Figure 1. SRM models simulating failure under direct shearing conditions (dashed line) showing the influence of the embedded DFN model on the modelled SRM results (modified from Elmo 2012).*

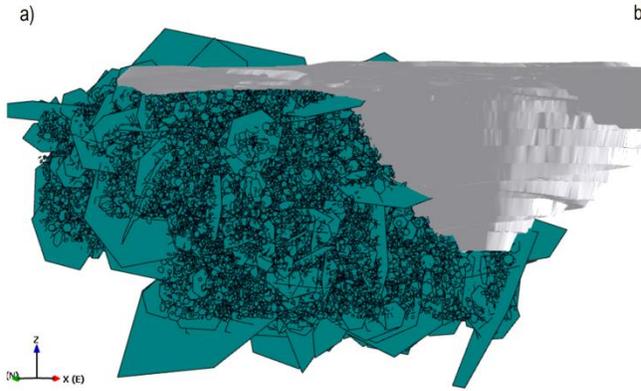
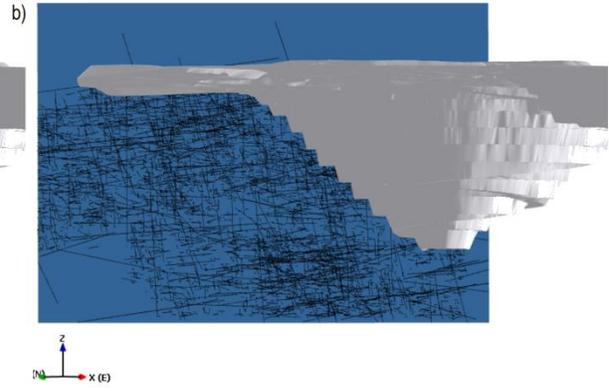

5_4_E — 45% Rock Bridge

2_1_E — 29% Rock Bridge

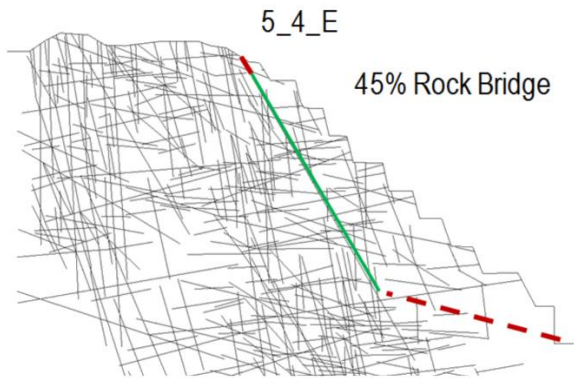
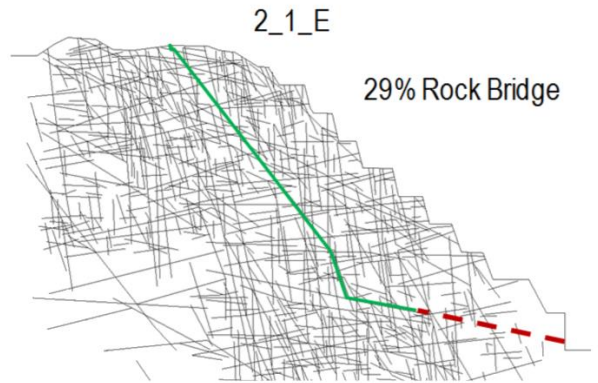

5_4_E1 — 25% Rock Bridge

2_1_E1 — 12% Rock Bridge

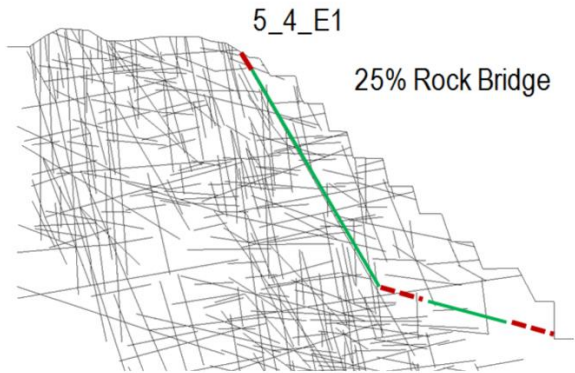
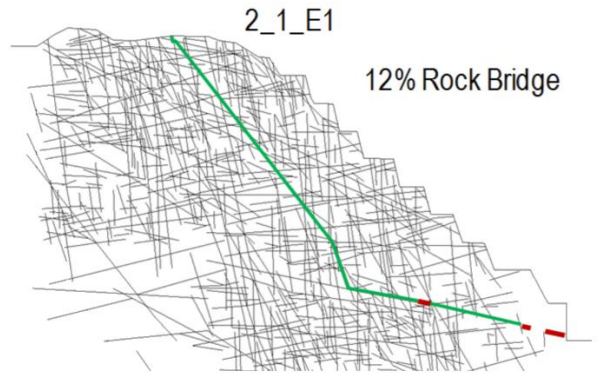

5_4_E2 — 5% Rock Bridge

2_1_E2 — 3% Rock Bridge

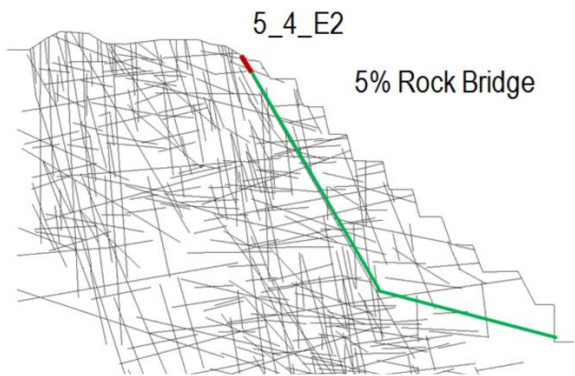
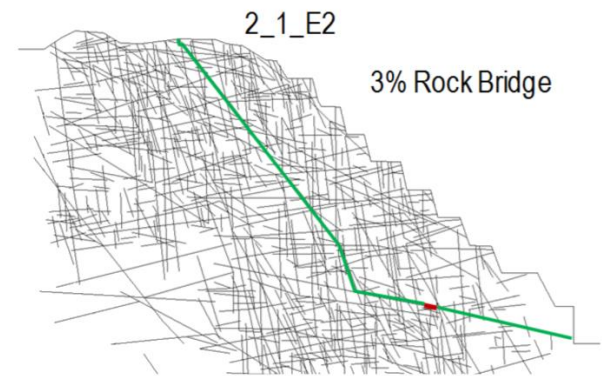

*Figure 2. (Top) Example of a three dimensional DFN model for: (a) a large rock slope problem; and (b) corresponding traces on a given two-dimensional section plane. (Bottom) Various models used in an integrated DFN-Geomechanical analysis of the same rock slope problem; modified from Elmo et al. 2011.*

## 2. DFN Characteristics

In the application of DFN modeling, the geometry of fractures in three-dimensional space can be modeled in numerous forms and shapes (Dershowitz et al. 2000); among these, flat polygons are the most versatile, sophisticated, realistic and also a recent trending shape representation. In practice, a single fracture is initially constructed by defining its size (i.e., its length along one or two axes for two- or three-dimensional fractures, respectively). Note that, the complexity in the shape of polygonal fractures can be addressed then by means of many specialized stochastic tools (Baecher 1983, Dershowitz & Einstein 1988). This includes a simple technique described in (Fadakar-A et al. 2011) and available in ADFNE (Fadakar-A 2017), which consists of generating four random values in the range of $[b, 1-b]$ where $b$ is a small value to prevent two connected nodes from overlying. The four values generate four nodes (corners) each on an edge of square. Following a full turn (clockwise or counterclockwise) which connects the four nodes a convex polygon is generated. The guide square can be replaced by a rectangle for more variant shapes in the resulting polygon.

Generally, the dominant characteristics of a fracture are its flatness and convexity (Zhang & Einstein 2000), in accordance to the fundamentals of DFN modeling (La-Pointe et al. 2002). Note that, the curviness observed in fractures in nature can be addressed by simplifying the curvy surface into several flat polygons. Such a simplification can be limited to a chosen resolution for a particular fracture network model. Generally, in order to structure a network, the generated fractures are then translated to a given three-dimensional domain by assigning random values to the fractures' centroid coordinates, i.e., $\{X, Y, Z\} \in \mathbb{R}$. In more advanced DFN modeling, a variety of stochastic methods including filtering techniques such as the rejection criterion (Diggle 2003) are utilized to generate the locations with respect to the desired point density (Diggle 2003, Baddeley et al. 2006). Finally, by incorporating the orientation information for all simulated fractures, a standard simplified procedure of DFN modeling is completed. The orientation information may include dip and dip-direction angles $\{0 \leq dip \leq 90, 0 \leq ddir < 360\}$ for a typical three-dimensional implementation.

In a reverse manner, a fracture network (e.g., based on sampled fracture data or a realization made by simulation) can efficiently and satisfactorily be characterized geometrically by extracting the location, orientation and size information of all fractures in the network regardless of the apparent complexity in the network (Figs.3 and 4). Therefore, intuitively speaking, any two DFN realizations can be compared by examining and quantifying any potential similarities between their corresponding characteristics. Basically, such an examination may include the study of the histogram and or rose-diagram (locus pattern) of the extracted orientation information, and also the histogram of the extracted size information for both fracture networks under comparison. An important note here is that, while histograms are generally useful for comparing orientation and size distributions in the study of fracture networks, the examination of location requires different approaches. Such a need is inherently associated with the spatial properties naturally linked to the distribution of points in space (Diggle 2003). Moreover, in order to compare two fracture networks for similarity, a single value outcome would be of more desire and extremely useful compared to a series of numbers (i.e., the frequency values from the histograms). Furthermore, a single value summarization would provide an opportunity to automate the comparison between sets of DFNs, and hence is particularly more helpful, for example, for selecting representative models for consequent processing and/or applications. The representativeness of a model here refers to the goals of a study, and therefore is entirely subjective to the project's aims. For example, if a DFN model is used to generate realizations as input models for intensive numerical investigations, the inputs must be at maximum distance in terms of similarity in order to encapsulate the uncertainty associated with the DFN model. In this case therefore, two less similar realizations are the best representatives. As a different scenario, say, one investigates an inherent secondary property of the DFN, for example the connectivity, which not always explicitly correlates with DFN input parameters. That is, the aim in the second case is to probe variation in the secondary property of realizations that are quite similar geometrically and spatially. For this scenario most similar realizations can be extracted by our proposed similarity metric as well.

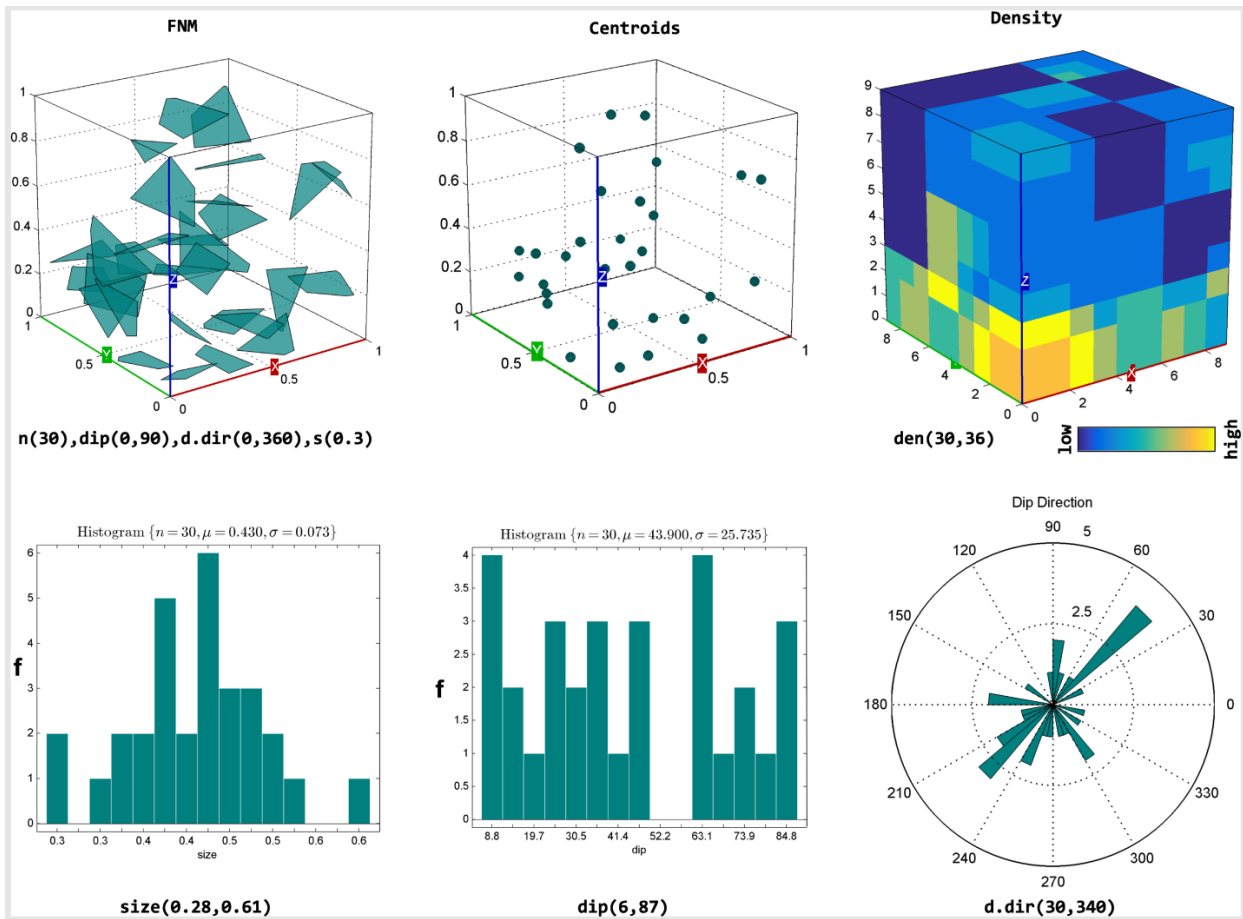

*Figure 3. Extraction of determinative characteristics from a simple DFN model (total of 30 fractures), including: location (centroids), fracture density, size, and dip and dip direction. The values between brackets report the minimum and maximum bounds for each parameter. See Fig.4 for complex case.*

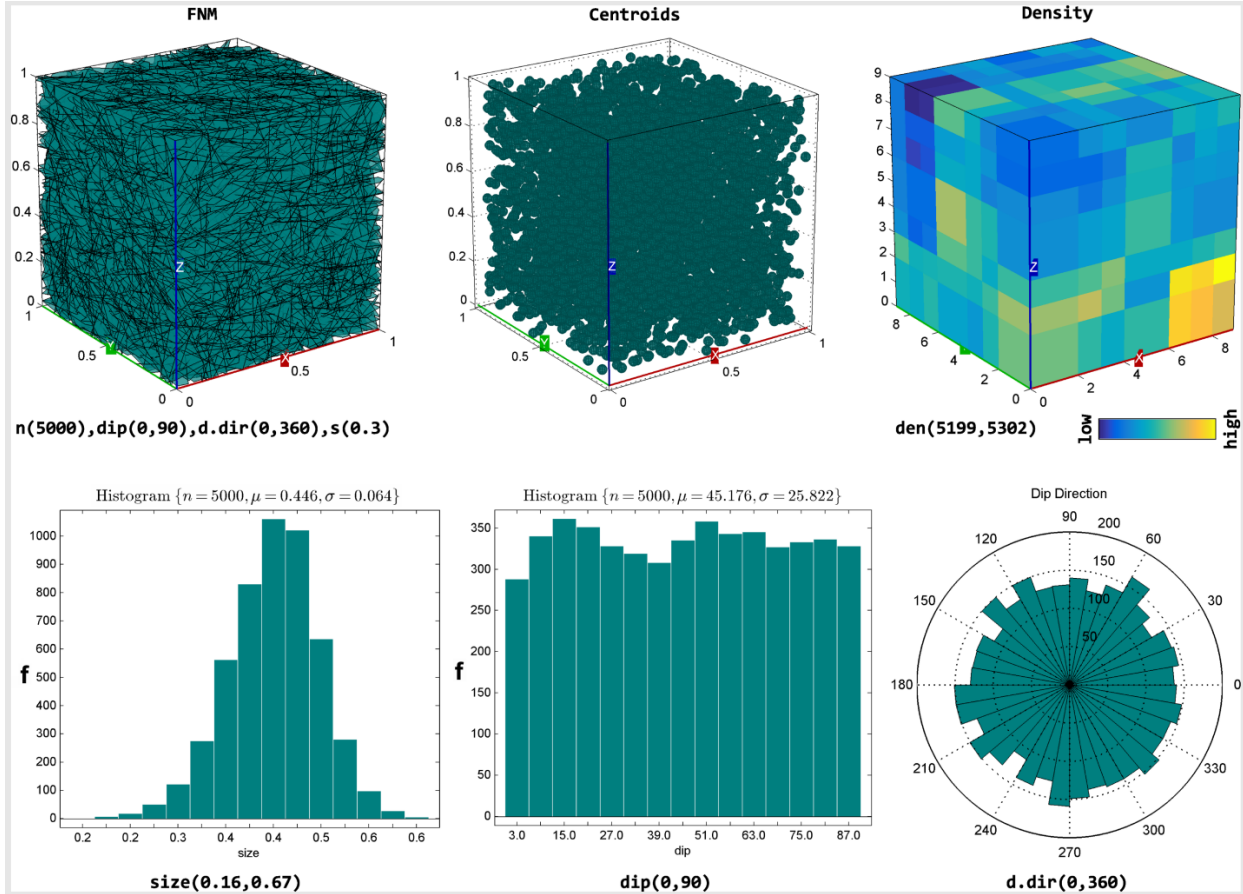

*Figure 2. Extraction of determinative characteristics from a complex DFN model (total of 5000 fractures), including: location (centroids), fracture density, size, and dip and dip direction. The values between brackets report the minimum and maximum bounds for each parameter.*

## 3. Similarity Analysis

Fundamentally, the histograms of data can satisfactorily be compared for similarities by means of the four standard metrics (OpenCV 2015) listed in Table 1. Strelkov (2008) discusses an optimization-based technique for histogram comparison which mimics the visual judgment of an expert. As can be seen from the formulas in Table 1 each metric summarizes (projects) the similarity between the two histograms into a bounded single value. For example, the result of a correlation metric varies between 0 and 1, corresponding to the minimum and maximum similarity between the two histograms under investigation. Table 1 also shows the boundary values known for the four metrics.

*Table 1. Limits for histogram similarity metrics*

| Method | Formula | Min Bound | Max Bound | Max Similarity |
|---|---|---|---|---|
| Correlation | $d(H_1, H_2) = \dfrac{\sum_i (H_1(i) - \bar{H}_1)(H_2(i) - \bar{H}_2)}{\sqrt{\sum_i (H_1(i) - \bar{H}_1)^2 \sum_i (H_2(i) - \bar{H}_2)^2}}$ | -1 | 1 | 1 |
| Chi-Square | $d(H_1, H_2) = \sum_i \dfrac{(H_1(i) - H_2(i))^2}{H_1(i)}$ | 0 | ? | 0 |

| | | | | | |
|---|---|---|---|---|---|
| Intersection | $d(H_1, H_2) = \sum_i \min(H_1(i), H_2(i))$ | | 0 | $\sum H_1$ | $\sum H_1$ |
| Bhattacharyya | $d(H_1, H_2) = \sqrt{1 - \frac{\sum_i \sqrt{H_1(i) \cdot H_2(i)}}{\sqrt{\bar{H}_1 \bar{H}_2 n^2}}}$ | | 0 | 1 | 0 |

\* ($H_1$ and $H_2$ are projected to $[0, 1]$)

Interestingly, the use of the similarity metric between two histograms appears efficient and helpful for that between two DFNs, particularly for orientation and size information. That is, if the two histograms show similar distribution parameters including first and higher order statistics and shape (namely mean, standard deviation etc.) for the size and/or orientations, the corresponding two fracture networks can be considered similar with respect to size and/or orientation. However, for fracture location(s) within the volume domain, the use of histograms cannot be applied directly since the locations are three-dimensional (i.e., not a vector). We propose here an intuitive quick solution by which one may proceed with an intermediate stage to transform the three dimensional location data to a vector. In the following sections, a complete multistage procedure is presented to address all of the mentioned issues above and to quantify the similarity between three-dimensional fracture networks. This involves a unique measure that takes into account all characteristics, information and their internal associations.

## 3.1. Framework for Similarity Analysis

The first step involves mapping the centroids of fractures into a three-dimensional density map by means of any point density estimator including sub-region counting (used in our implementation), Kernel Density Estimator (KDE; Duong 2004), etc. Then an Especial Spatial Dimension Reduction (ESDR) technique (described below) is used to map three-dimensional density information onto a one-dimensional curve (generating a vector of the histogram frequency values). A simple procedure for this step would involve box sampling sub-stages in which the entire domain is fully divided into small three-dimensional cubes (sub-domains or sub-regions). Every cube is then assigned a single density value determined by calculating the number of enclosed fracture centroids inside the cube's volume (i.e., counting method, see also density map in Figs.3 and 4). The resulting three-dimensional matrix of density values can then be flattened by visiting all $i$, $j$ and $k$ indices (e.g., ESDR), producing a one-dimensional curve (a vector), which is basically a histogram of frequency values for points in a cube. As a result, all four similarity metrics for histograms can now be used for conducting comparison between fracture networks for sizes, orientations and locations. The suggested steps for the full DFN similarity algorithm are summarized below:

1. Determine centroids of fracture in the DFNs → $cts_1 = \{\bar{X}, \bar{Y}, \bar{Z}\}$ and $cts_2 = \{\bar{X}, \bar{Y}, \bar{Z}\}$
2. Compute density of $cts_1$ and $cts_2$ → $den_1 = f(x \in V_1)$ and $den_2 = f(x \in V_2)$
3. Find metrics of differences between $den_1$ and $den_2$ → $(c_0, h_0, i_0, b_0)$
4. Calculate histogram of lengths for fractures in the DFNs → $szs_1$ and $szs_2$
5. Find metrics of differences between $szs_1$ and $szs_2$ → $(c_1, h_1, i_1, b_1)$
6. Calculate histogram of dips for fractures in the DFNs → $ds_1$ and $ds_2$
7. Find metrics of differences between $ds_1$ and $ds_2$ → $(c_2, h_2, i_2, b_2)$
8. Find rose of dip directions for fractures in the DFNs → $dds_1$ and $dds_2$
9. Find metrics of differences between $dds_1$ and $dds_2$ → $(c_3, h_3, i_3, b_3)$
10. Report $\{\bar{c}, \bar{h}, \bar{\iota}, \bar{b}\}$, where e.g., $\bar{c} = 0.25 * (c_0 + c_1 + c_2 + c_3)$

Note that in step 10 above, the 0.25 multiplier is to apply mathematical average to the four metrics; i.e., no weighting is applied. In more advanced uses, however, one may define different weights for density ($c_0$, $c_2$) as well as for size ($c_1$, $c_3$). The higher the weight, the more emphasis is imposed on the difference; therefore according to the goals of the similarity analysis, one may apply a desired weighting system. Other interpretations would also be beneficial such as taking the maximum or minimum of the four metrics. Ultimately if there was not a legitimate reason or preference to differentiate the metrics by weighting factors, the suggested simple mathematical averaging would suffice.

The described solution works as expected; however, further in-depth observations show that the above method is lacking the incorporation of the association between characteristics in the similarity quantification. Hence, it would be vulnerable to introduce erroneous (and or biased) results due to the missing linkage. This issue is demonstrated in Fig.5 in which several steps in the DFN similarity algorithm are shown (from left to right) including the extraction of location, size and orientation (dip and dip-direction angles) information. Density maps are also shown. The first row in Fig.5 demonstrates a reference realization from a Fracture Network Model (FNM). The second row shows another realization (from the same model), while the third row is the same as the reference realization but affected by a developed exchange function. To utilize our investigations we developed an exchange algorithm: the location of a randomly chosen pair of fractures is exchanged. Hence, by application of the exchange algorithm, only the links between the location information and other fracture information are altered. That is, one should expect to have the same histogram (locus pattern) for orientations and sizes, and the same density map for locations for both the original and exchanged fracture networks, although interlinks would be different. The steps in the exchange algorithm developed are:

1. Take a fracture and find its boundary box → $bbox_1$
2. Take another fracture and find its boundary box → $bbox_2$
3. Check if both $bbox_1$ and $bbox_2$ are exchangeable i.e., not crossing region boundaries
4. If so, exchange their locations, otherwise repeat from 2

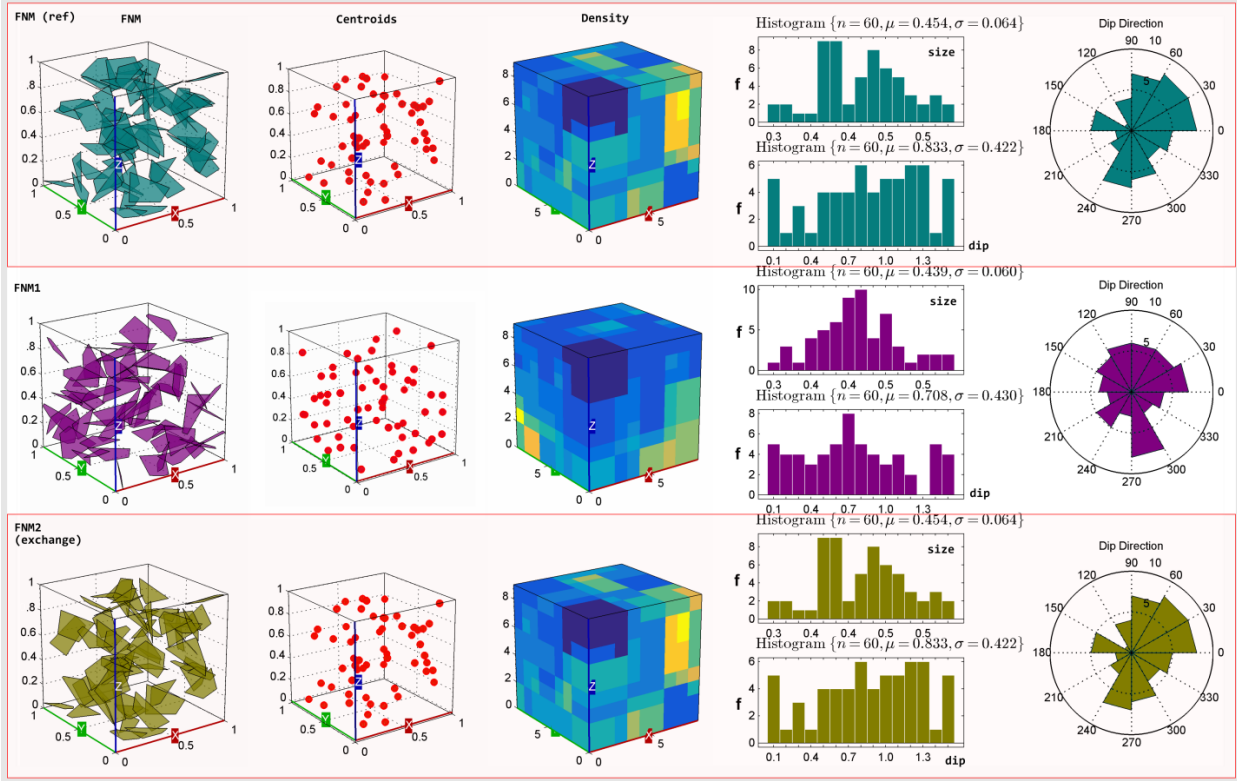

*Figure 5. Steps in the DFN similarity algorithm (from left to right) including the extraction of centroid location, centroid density, fracture size and fracture orientation. The first row depicts a reference realization, which is compared with a realization from the same model (second row), as well as the reference realization affected by a developed exchange function (third row). Note that, due to the missing linkage between the location and orientation information, the top and bottom rows shows exact same DFN characteristics although they are not the same.*

A series of example applications of the exchange algorithm is shown in Fig.6. As demonstrated by increasing the rate of exchange, we would expect declination in the similarities between the manipulated realization (i.e., affected by exchange function) and the reference fracture network due to the probability associated with random perturbations. That is, there is little chance a fracture to be exchanged by a quite similar fracture. The effectiveness and suitability of the proposed exchange algorithm for our study was investigated and is confirmed as shown in Figs.6 and 7. In Fig.7, the result of a simulation (100 tests per case) for varying exchange rate and the corresponding similarity values are (3|0.9732, 6|0.9461, 12|0.8968, 24|0.8168, 48|0.8105).

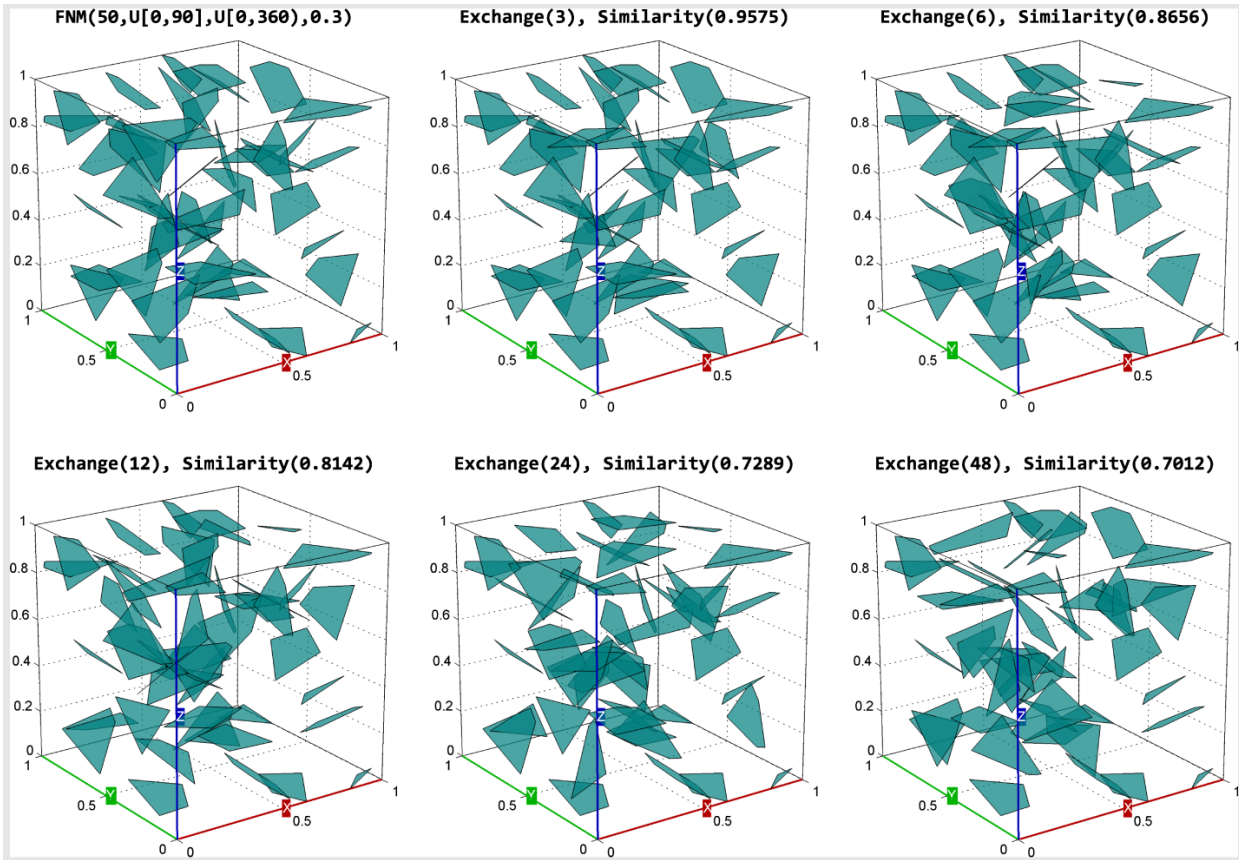

*Figure 6. Effectiveness of Exchange algorithm on the Similarity. The similarity decreases as the rate of exchange increases due to the probability of perturbation.*

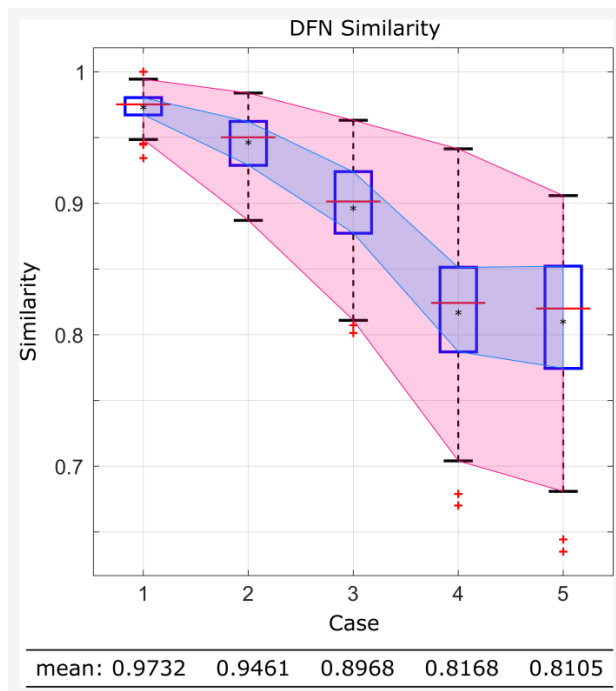

*Figure 7. DFN Similarity evaluation on varying exchange number 3, 6, 12, 24 and 48 (cases 1 to 5, respectively) over the reference DFN model; total number of tests per case: 100.*

Briefly, the application of the exchange function provided us an opportunity to study the association (linkage) between fractures characteristics. Returning to the similarity assessment, one solution for studying the missing linkage between location and orientation information would be to associate the resulting location density to the local orientation information. That is, any density cell exposed as a single value on the final histogram (refer to previous sections for details) now reflects the variation in the orientation, as well. The full algorithm of this proposal is presented below:

1. Determine centroids of fracture in the DFNs → $cts_1 = \{\bar{X}, \bar{Y}, \bar{Z}\}$ and $cts_2 = \{\bar{X}, \bar{Y}, \bar{Z}\}$
2. Categorize $cts_1$ and $cts_2$ based on dip angle → $cts_{11}^*$ and $cts_{21}^*$
3. Compute density of $cts_{11}^*$ and $cts_{21}^*$ → $den_{11} = f(x \in V_1)$ and $den_{21} = f(x \in V_2)$
4. Categorize $cts_1$ and $cts_2$ based on dip direction angle → $cts_{12}^*$ and $cts_{22}^*$
5. Compute density of $cts_{12}^*$ and $cts_{22}^*$ → $den_{12} = f(x \in V_1)$ and $den_{22} = f(x \in V_2)$
6. Find metrics of differences between $den_{11}$ and $den_{21}$ → $(c_0, h_0, i_0, b_0)$
7. Find metrics of differences between $den_{12}$ and $den_{22}$ → $(c_1, h_1, i_1, b_1)$
8. Calculate histogram of lengths for fractures in the DFNs → $szs_1$ and $szs_2$
9. Find metrics of differences between $szs_1$ and $szs_2$ → $(c_2, h_2, i_2, b_2)$
10. Report $\{\bar{c}, \bar{h}, \bar{i}, \bar{b}\}$, where e.g., $\bar{c} = (c_0 + c_1 + c_2)/3$

By conducting steps 2 and 4 above, the orientation information is categorized into four sets, i.e., two sets for every dip and dip-direction information. Computing densities in steps 3 and 5 link the location information to the orientation information. The resulting values are treated as histograms that together with size histograms undergo the application of similarity metrics in Table 1. Finally, the DFN similarity metric is calculated as the average of the three resulting indices reported in step 10. In the presented algorithm, one has flexibility to categorize the input into even more groups, for example, by setting the orientation cuts at 45 degree intervals resulting in 8 categories. Increasing the number of categories would likely lead to more detailed comparisons, up to a limit. However, it would also increase the cost and complexity of computation. Again, depending on the goals of the similarity analysis, one may utilize geostatistical, spatial and/or directional statistical techniques in order to justify the number of categories; the framework proposed here remains unchanged.

The proposed similarity measure, as shown, is capable of quantifying any difference between a pair of realizations which is otherwise hidden. For example, a very heterogeneous fracture network if its rotated relative to one axis would result in the same permeability and bulk modulus but a low value of similarity. This is critical specifically for applications where preferred pathways through fracture networks are of interest. That is, our similarity measure can quantify how different the two fracture networks might be.

## 4. Sensitivity Analysis for the Proposed Metric

To further develop our study we tested the sensitivity of the proposed DFN similarity metric to the variation in characteristics of fractures. To do so, we conducted a simple Monte Carlo analysis to investigate the trend of variation in the resulting similarity metric due to variation in several input parameters: the centroid locations, fracture orientations, and fracture sizes, based on three-dimensional DFN modeling. Further details on Monte Carlo methods can be found in Rubinstein & Kroese (2008).

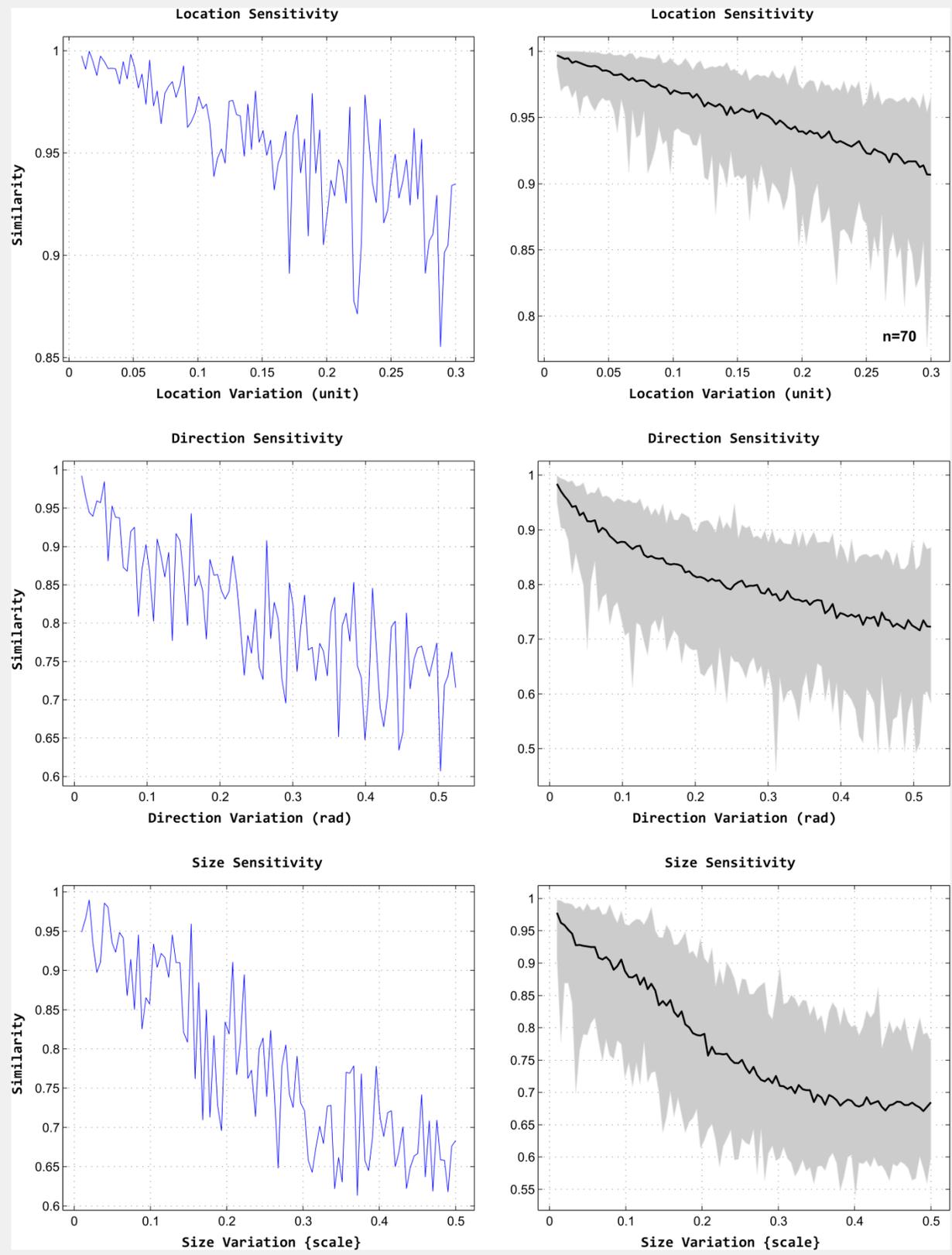

*Figure 8. Sensitivity Analysis of Similarity measure to the variation in DFN model parameters: centroid location, fracture direction and fracture size. The average curves in the right column are*

*based on 70 independent simulations. In each simulation a 100 stepped variation in the parameters was applied.*

For centroid locations, the variation range was [0.01,0.3] units. The orientation information underwent a change between 0.5 and 30 degrees. A range of [0.01,0.5] units was used to modify the fracture size information. In Fig.8 the responses of the proposed measure to the variation in the model parameters including location ({X,Y,Z}), direction ({dip,dip-direction}) and size (L), are shown. Interestingly, the 70 independent simulations conducted resulted in nicely shaped averaged curves that consistently demonstrate declination in the similarity by increasing variation of each parameter for any two compared fracture networks. Therefore our proposed three-dimensional similarity measure could fully capture different changes quite well and hence suggests a fairly comprehensive coverage of variation in the model parameters.

## 5. Comparison between three-dimensional and two-dimensional Similarity Measures

Fractures and fracture networks in rock mass are three-dimensional, as are their associated characteristics. Nevertheless, the majority of reported studies in the literature including concepts, techniques, methods and frameworks focus on two-dimensional space mainly due to the complexity of the problem of study; examples are works by de-Dreuzy et al. (2001), Parashar & Reeves (2012), Fadakar-A et al. (2013b) and Lei et al. (2014). However, the use of two-dimensional solutions likely results in biased and/or erroneous findings, that is, the accuracy, reliability and representativeness of the solutions will be highly doubtful. As a workaround, if there was a mathematically robust and meaningful way to calibrate the results of a two-dimensional solution with respect to its three-dimensional solution, then careful and supervised use of two-dimensional solutions might be tolerated and so justified. Searching for such strong relationships is a hard problem in itself. A relatively simple approach is to utilize Monte Carlo simulations (Robert & Casella 1999) to make judgments on the resulting behaviors of three-dimensional and two-dimensional solutions more reliable. Doing so, if there was a good correlation between the results then perhaps appropriate coefficients can be determined for the justification and adjustment of two-dimensional solutions. In the following sections we applied this technique (i.e., Monte Carlo modeling) to the similarity problem.

In our implementation of two-dimensional similarity applied to three-dimensional fracture networks, a set of three groups of planes (i.e., flat two-dimensional profiles) aligned to the three global axes ({X,Y,Z}) are set to intersect every realization of a three-dimensional DFN. The full procedure is shown in Fig.9. This produces a number of trace-planes which are then compared with the corresponding trace-plane on the reference DFN. The method to compare the two planes of traces (i.e., two-dimensional fracture networks) was introduced in Fadakar-A et al. (2013a), which basically applies the distance weighting cosine similarity approach (see the reference for details and discussions). Figures 10 and 11 demonstrate two example cases of the two-dimensional method. As shown, the maximum similarity value (i.e., 1) between the two fracture networks is achieved when one overlays the other one fully and precisely, whereas the minimum similarity value (i.e., 0) is found when the two are perpendicular to each other. In Fig.11, two example realizations are shown that resulted in different similarities to a reference fracture network. The method works well for any size of and complexity in fracture networks.

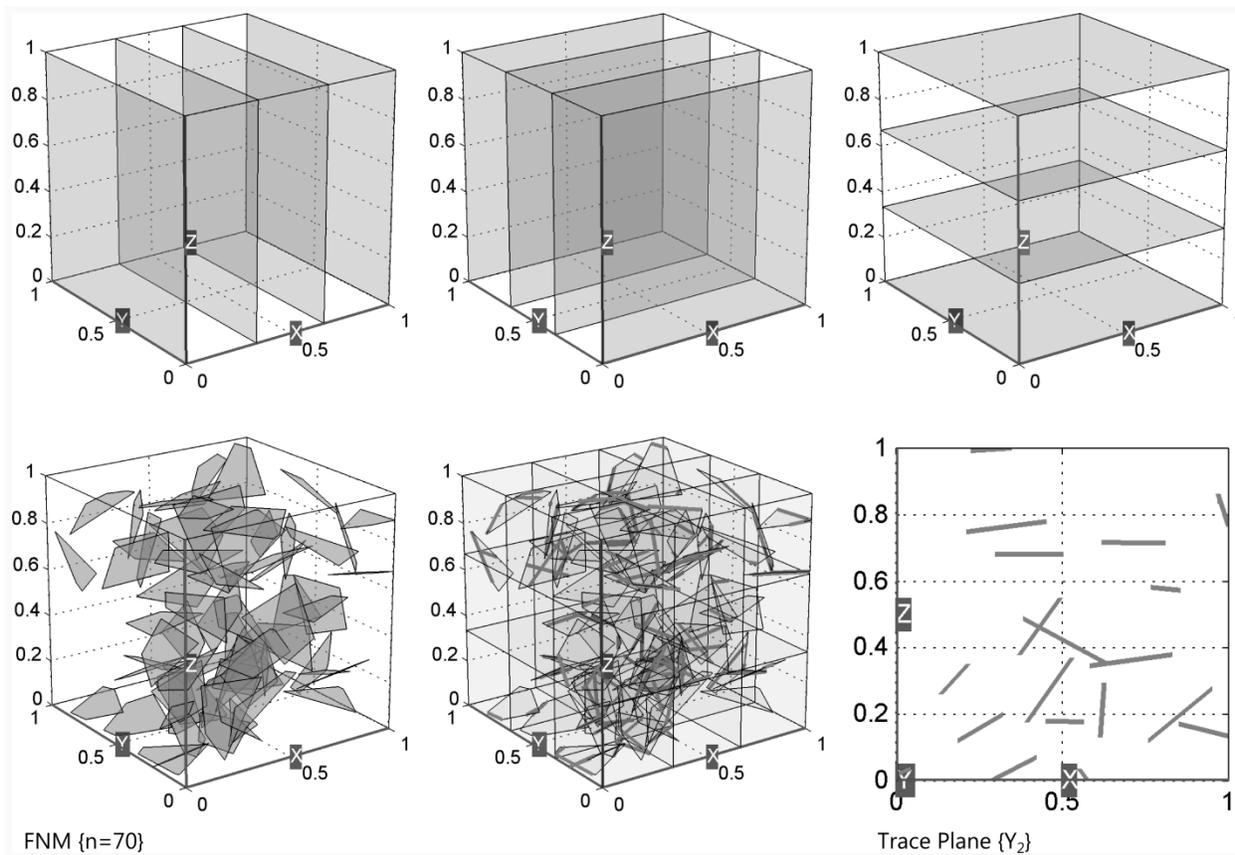

*Figure 9. Stages of plane sampling on three-dimensional fracture networks for evaluation of two-dimensional similarity between trace planes (top-left to bottom-right).*

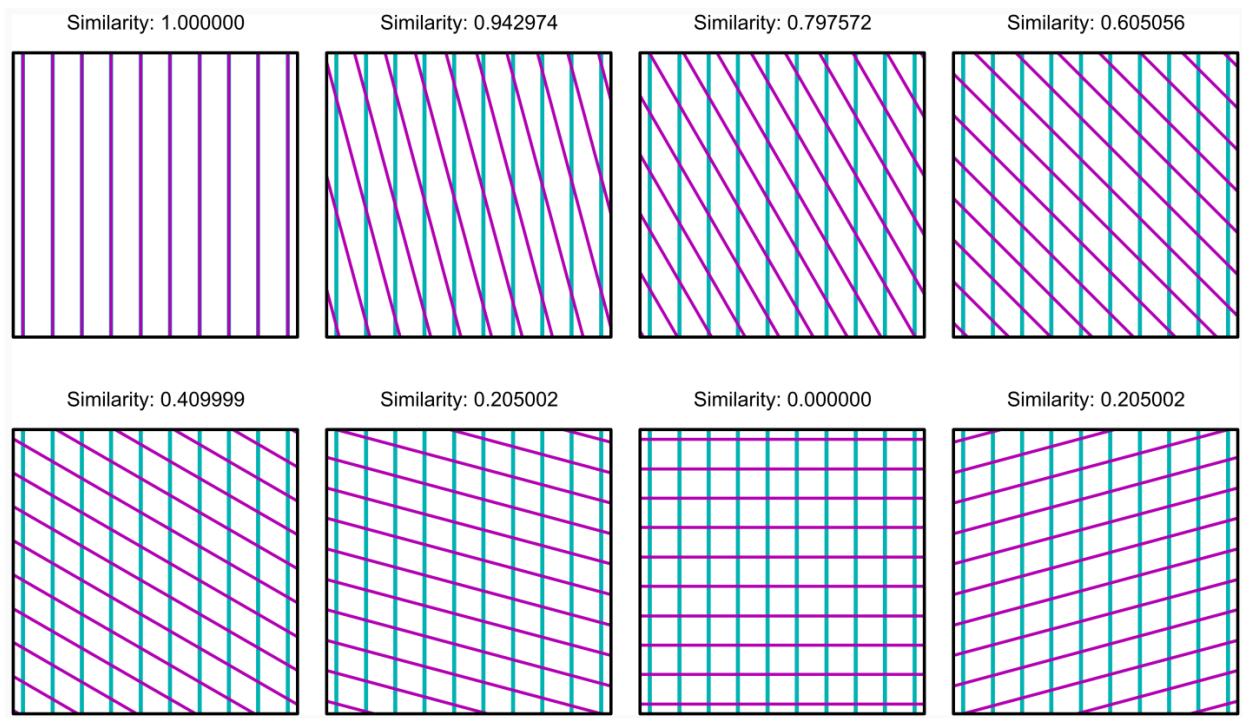

*Figure 10. Similarity values obtained for two sets of two-dimensional lines. Maximum and minimum similarities correspond to the identical (top-left: 1) and perpendicular cases (second bottom-right: 0).*

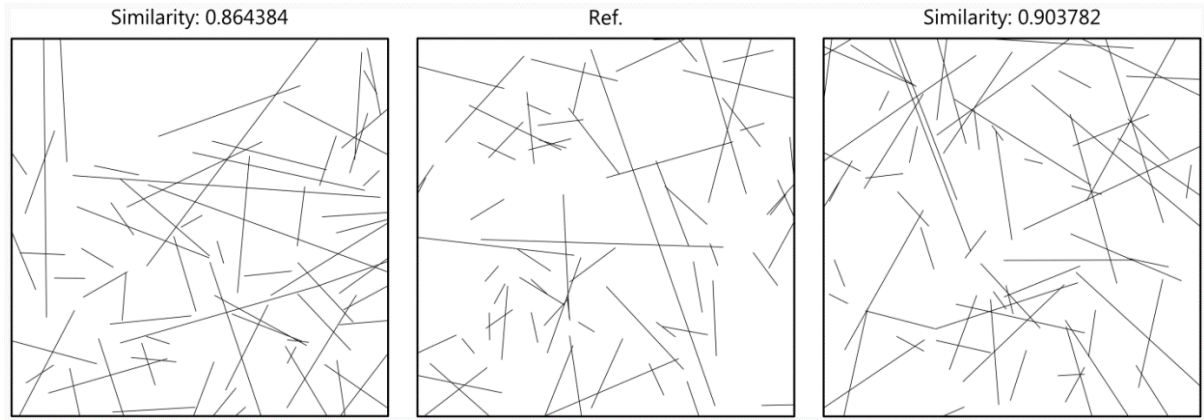

*Figure 11. Similarities evaluated on two two-dimensional DFN realizations against a reference fracture network.*

A comparison between similarity values obtained by the proposed three-dimensional method presented earlier and the two-dimensional method described in Fadakar-A et al. (2013a) was conducted. The settings for the fracture network model were as follows. A total number of 101 realizations were generated with each consisting of 70 fractures with orientation information ($\overline{dip} = m_{dip} = 45$, $variation = \kappa_{dip} = 5$ and $\overline{dip\_direction} = m_{ddir} = 180$, $variation = \kappa_{ddir} = 5$). Locations followed a uniform random distribution, and polygonal shapes (and sizes) were determined by means of the method described in Fadakar-A et al. (2011). The simulation was bounded in a cube of size 1x1x1 units; hence, a three-dimensional clipping procedure was in place (Diggle 2003). One randomly chosen realization from the simulated 101 realizations was considered as the reference fracture network. The comparison of the results is shown in Fig.12. The results indicate that there is no correlation whatsoever between the similarity values derived from the two- and three- dimensional similarity measures.

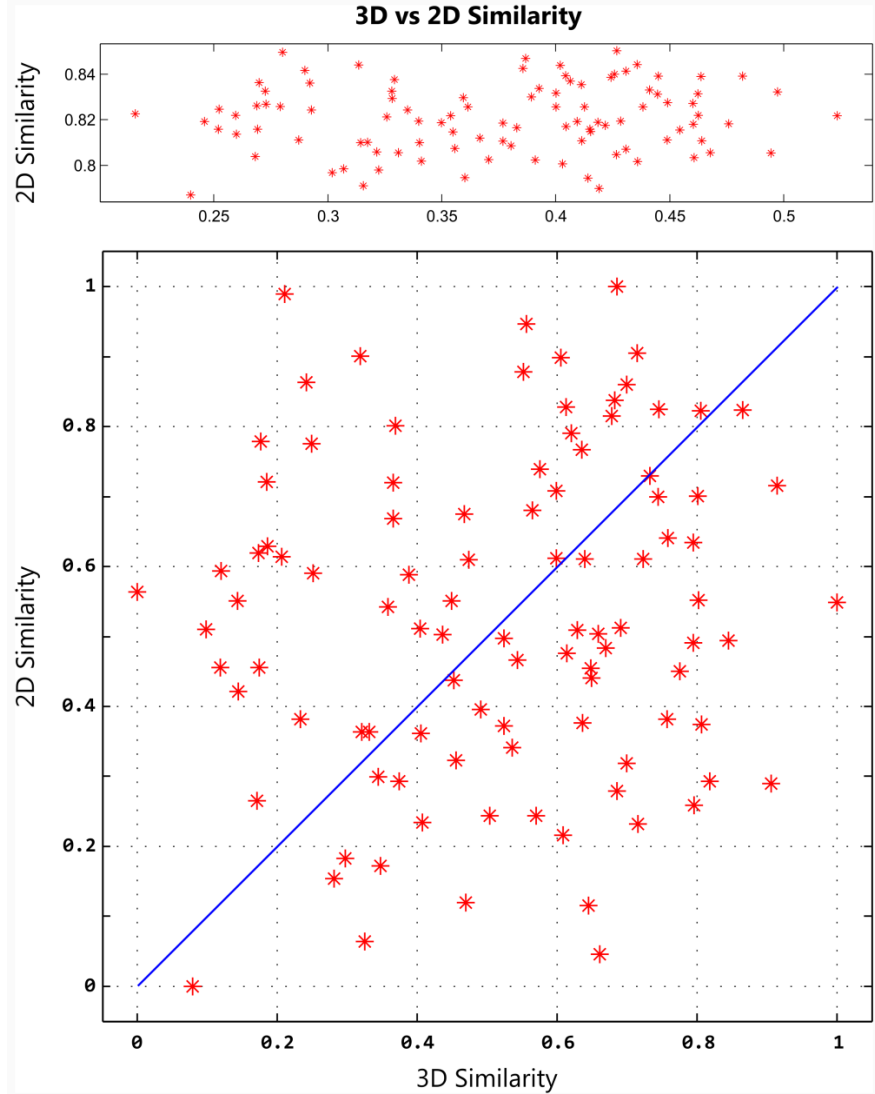

*Figure 12. Comparison of three- and two-dimensional similarities showing no correlation. In the bottom chart the similarity values are projected to [0,1] for better comparison, i.e., around 45 degree diagonal reference line.*

Further investigation was made by means of conducting a simulation of DFN under settings ($[\text{sim}] = 10, [\text{fnm}] = 70, \{dip\left(\frac{\pi}{4}, \kappa = 0.5\right), d.dir(\pi, \kappa = 0.5), size_{max} = 0.3\}$). For all pairs of realizations the similarity indices were computed producing two similarity matrices. The aim was to find the most similar two realizations among all pairs. These are pairs (1,3) and (6,8) as shown in Fig.14. Note that we combined the two matrices in a one square matrix for compact and quick comparison. As shown, realizations 1 and 3 found through the two-dimensional similarity metric and realization 6 and 8 found by the three-dimensional similarity metric are completely different. The results show significant differences in the overall pattern and distribution of similarity values (color coded), further indicating the limitations of two-dimensional similarity based analyses.

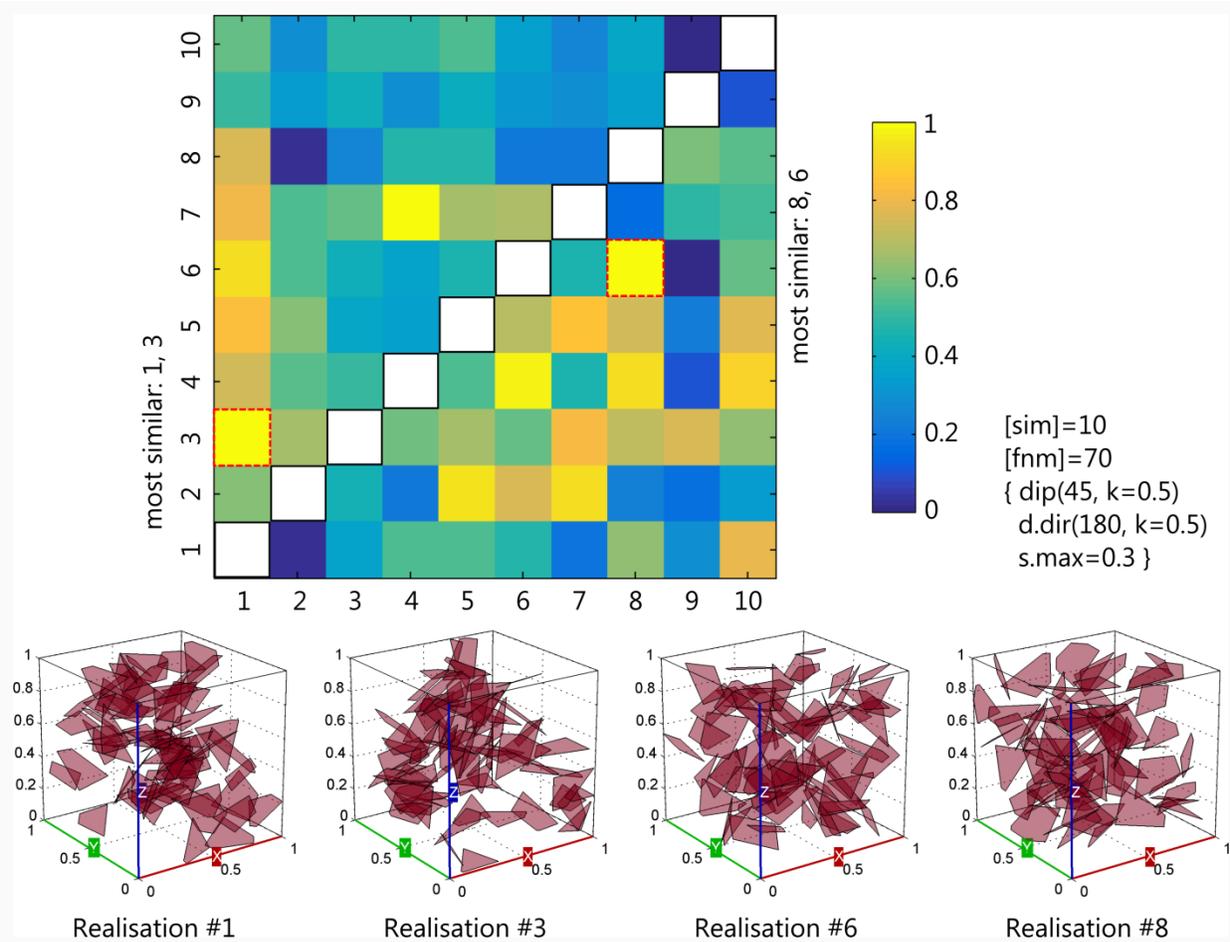

*Figure 14. Similarity matrices constructed based on 10 realizations (warm colors indicate similarity). The most similar pair obtained by two-dimensional analysis is pair (1,3) and that for three-dimensional similarity is pair (6,8).*

## 6. Conclusions

Three-dimensional DFN models better represent the reality of rock mass fractures in terms of the complexity of geometry and spatial characteristics. The similarity between DFN realizations was studied in this paper. Proposed algorithms made possible the robust assessment of the similarity for three-dimensional fracture networks covering all variations in their characteristics. A similarity metric was developed and shown to be a practical, effective and useful indicator, as demonstrated through the example of a geomechanical model of a large open pit slope. A sensitivity analysis was conducted by means of extensive Monte Carlo simulations and showed that the metric is responsive to variations in locations, orientations and sizes within the fracture networks.

As an intuitive application of the proposed method, a comparative study was made between the proposed and a previously developed two-dimensional similarity metric. The results showed that the similarity metrics do not correlate. This outcome is very important for validating analyses based on two-dimensional similarity studies. It is therefore recommended here that three-dimensional similarity be used in order to select similar or dissimilar DFN models/realizations for advanced engineering analyses and designs.